\input harvmac
\writedefs
\sequentialequations
\overfullrule=0pt
\def\comment#1{}

\def\+{^\dagger}

\def\e{e}

\def \Q {{\hat Q}}
\def \P {{\hat P}}
\def \q {{\hat q}}
\def \Qk {{\hat Q_K}}

\def \del {\partial}

\def \a {\alpha}

\def \chi {\chi}

\def\np {{\it  Nucl. Phys. }}
\def \pl {{\it  Phys. Lett. }}

\def \prl {{\it  Phys. Rev. Lett. }}
\def \pr  {{\it Phys. Rev. }}

\def \Q {{\hat Q}}
\def \P {{\hat P}}
\def \q {{\hat q}}

\def \del {\partial}

\def \a {\alpha}

\def \chi {\chi}

\def\np {{\it  Nucl. Phys. }}
\def \pl {{\it  Phys. Lett. }}

\def \prl {{\it  Phys. Rev. Lett. }}
\def \pr  {{\it Phys. Rev. }}


\def\comment#1{}
\def\fixit#1{}



\def\sqr#1#2{{\vcenter{\vbox{\hrule height.#2pt
         \hbox{\vrule width.#2pt height#1pt \kern#1pt
            \vrule width.#2pt}
         \hrule height.#2pt}}}}





\def\+{^\dagger}

\def\overleftrightarrow#1{\vbox{\ialign{##\crcr
     \leftrightarrow\crcr\noalign{\kern-0pt\nointerlineskip}
     $\hfil\displaystyle{#1}\hfil$\crcr}}}


\Title{
 \vbox{\baselineskip10pt
  \hbox{PUPT-1692}
  \hbox{hep-th/9703216}
 }
}
{
 \vbox{
  \centerline{ Testing Effective String Models of Black Holes}
  \vskip 0.1 truein
  \centerline{with Fixed Scalars }
 }
}
\vskip -25 true pt

\centerline{
 Michael Krasnitz and Igor R.~Klebanov 
 }
\centerline{\it Joseph Henry Laboratories, 
Princeton University, Princeton, NJ  08544}

\bigskip
\centerline {\bf Abstract}
\bigskip

We solve the problem of mixing between the fixed scalar and metric
fluctuations. First, we derive 
the decoupled fixed scalar equation for the four-dimensional 
black hole with two different
charges. We proceed to the five-dimensional black hole 
with different electric (1-brane) and magnetic (5-brane) charges,
and derive two decoupled equations 
satisfied by appropriate mixtures of the
original fixed scalar fields. The resulting greybody factors are 
proportional to those that follow from coupling to dimension (2,2)
operators on the effective string. In general, however, the
string action also contains couplings to
chiral operators of dimension (1,3) and (3,1), which cause
disagreements with the semiclassical absorption cross-sections.
Implications of this for the effective string models are discussed.

\Date{March 1997}

\noblackbox
\baselineskip 14pt plus 1pt minus 1pt


\def\jref{}  

\lref\ATT{A.A.~Tseytlin, {\it Mod.~Phys.~Lett.}~A11 (1996) 689,
hep-th/9601177.}

\lref\Ark{A. Tseytlin, {\it Nucl. ~Phys.}~B469 (1996) 51, hep-th/9602064.}

\lref\CGKT{
C.G.~Callan, Jr., S.S.~Gubser, I.R.~Klebanov and A.A.~Tseytlin,
{\it Nucl. ~Phys.}~B489 (1997) 65, hep-th/9610172;
I.R. Klebanov and M. Krasnitz, {\it Phys.~Rev.}~D55 (1997) 3250, 
hep-th/9612051.}

\lref\CTT{M. Cveti\v c and  A.A.  Tseytlin, 
\pl B366 (1996) 95, hep-th/9510097; \pr D53 (1996) 5619, 
hep-th/9512031.}

\lref\CT{M.~Cveti\v c and A.A.~Tseytlin, {\it Nucl.~Phys.}~B478 (1996)
181, hep-th/9606033.}

\lref\CYY{M.~Cveti\v c and D.~Youm, {\it Nucl.~Phys.}~B476 (1996) 118,
hep-th/9603100\jref.}

\lref\CY{M.~Cveti\v c and D.~Youm, {\it Phys.~Rev.}~D53 (1996) 584,
hep-th/9507090; Contribution to `Strings 95', hep-th/9508058.}

\lref\Duff{M.J. Duff, H. Lu, C.N. Pope
Phys. Lett. B382 (1996) 73, hep-th/9604052.}

\lref\GKtwo{S.S.~Gubser and I.R.~Klebanov, {\it Phys. Rev. Lett.} 77
(1996) 4491, hep-th/9609076.}

\lref\GK{S.S.~Gubser and I.R.~Klebanov, 
{\it Nucl. Phys.} B482 (1996) 173, hep-th/9608108.}

\lref\GR{I.S.~Gradshteyn and I.M.~Ryzhik, {\it Table of Integrals,
Series, and Products}, Fifth Edition, A.~Jeffrey, ed. (Academic Press:
San Diego, 1994).}

\lref\HK{A. Hashimoto and I.R. Klebanov, 
\pl B381 (1996) 437, hep-th/9604065.}

\lref\HM{G.~Horowitz and A.~Strominger, {\it Phys.~Rev.~Lett.}~77
(1996) 2368, hep-th/9602051.}

\lref\HP{G. Horowitz and J. Polchinski, hep-th/9612146.}

\lref\JP{J. Polchinski, \prl 75 (1995) 4724, hep-th/9510017.}

\lref\KT{I.R. Klebanov and A.A. Tseytlin, \np B475 (1996) 179, 
hep-th/9604166.}

\lref\LS{L. Susskind, hep-th/9309145.}

\lref\krt{ I.R.~Klebanov, A.~Rajaraman and A.A.~Tseytlin,
hep-th/9704112.}

\lref\LWW{F.~Larsen and F.~Wilczek, {\it Phys.~Lett.}~B375 (1996) 37,
hep-th/9511064; hep-th/9609084\jref.}

\lref\Unruh{W.G.~Unruh, {\it Phys.~Rev.}~D14 (1976) 3251.}

\lref\Wald{R.M.~Wald, {\it General Relativity} (Chicago: The
University of Chicago Press, 1984).}

\lref\at{A.A.~Tseytlin, {\it Nucl.~Phys.}~B475 (1996) 149,
hep-th/9604035.}

\lref\cm{C.G.~Callan and J.M.~Maldacena, {\it Nucl.~Phys.}~B472 (1996)
591, hep-th/9602043.}

\lref\dgm{S.~Das, G.~Gibbons and S.~Mathur, \prl 78 (1997) 417, hep-th/9609052}

\lref\dmII{S. Das and S.D. Mathur, {\it Nucl. ~Phys.}~B482 (1996) 153,  
hep-th/9607149.}

\lref\dmI{S.R.~Das and S.D.~Mathur, {\it Phys.~Lett.}~B375 (1996) 103,
hep-th/9601152.}

\lref\dmw{A.~Dhar, G.~Mandal and S.~R.~Wadia, {\it Phys.~Lett.}~B388 (1996) 51, Tata preprint TIFR-TH-96/26, hep-th/9605234.}

\lref\dm{S.R.~Das and S.D.~Mathur, {\it Nucl.~Phys.}~B478 (1996) 561,
hep-th/9606185; hep-th/9607149\jref.}

\lref\dowk{F. Dowker, D. Kastor and J. Traschen, hep-th/9702109.}

\lref\fkk{S.~Ferrara and R.~Kallosh, {\it Phys.~Rev.}~D54 (1996) 1514,
hep-th/9602136; {\it Phys.~Rev.}~D54 (1996) 1525, hep-th/9603090;
S.~Ferrara, R.~Kallosh, A.~Strominger, {\it Phys.~Rev.}~D{52} (1995)
5412, hep-th/9508072.}

\lref\gibb{G.~Gibbons, {\it Nucl.~Phys.}~{B207} (1982) 337;
P.~Breitenlohner, D.~Maison and G.~Gibbons, {\it
Commun.~Math.~Phys.}~120 (1988) 295.}

\lref\gkk{G.~Gibbons, R.~Kallosh and B.~Kol, \prl77 (1996) 4992, 
hep-th/9607108}

\lref\gkt{J.P.~Gauntlett, D.~Kastor and J.~Traschen, {\it
Nucl.~Phys.}~B478 (1996) 544, hep-th/9604179.}

\lref\gunp{S.S.~Gubser, November~1996, unpublished notes.}

\lref\hawk{S. Hawking and M. Taylor-Robinson, hep-th/9702045 .}

\lref\hmf{{\it Handbook of Mathematical Functions}, M.~Abramowitz and
I.A.~Stegun, eds. (US Government Printing Office, Washington, DC,
1964) 538ff.}

\lref\hms{G.~Horowitz, J.~Maldacena and A.~Strominger, {\it
Phys.~Lett.}~B383 (1996) 151, hep-th/9603109.}

\lref\hrs{E.~Halyo, B.~Kol, A.~Rajaraman and L.~Susskind,
hep-th/9609075\jref; E.~Halyo, hep-th/9610068\jref.}

\lref\hs{G.~Horowitz  and A.~Strominger, Nucl. Phys. B360 (1991) 197.}

\lref\jpTASI{J.~Polchinski, hep-th/9611050\jref.}

\lref\juanI{J.~Maldacena, Rutgers preprint RU-96-102,
hep-th/9611125\jref.}

\lref\juan{J.~Maldacena, {\it Nucl.~Phys.}~B477 (1996) 168,
hep-th/9605016.}

\lref\kaaa{R.~Kallosh, A.~Linde, T.~Ort\'in, A.~Peet and A.~Van
Proeyen, {\it Phys.~Rev.}~D{46} (1992) 5278, hep-th/9205027.}

\lref\km{I.R.~Klebanov and S.D.~Mathur, Princeton preprint PUPT-1679,
MIT preprint MIT-CTP-2610, hep-th/9701187.\jref}

\lref\kr{B.~Kol and A.~Rajaraman, Stanford preprint SU-ITP-96-38,
SLAC-PUB-7262, hep-th/9608126\jref.}

\lref\ktI{I.R.~Klebanov and A.A.~Tseytlin, Princeton preprint
PUPT-1639, {\it Nucl.~Phys.}~B479 (1996) 319, hep-th/9607107.}

\lref\kt{I.R. Klebanov and A.A. Tseytlin, \np B475 (1996) 165, 
hep-th/9604089.}

\lref\lu{J.X.~Lu, {\it Phys.~Lett.}~B313 (1993) 29, hep-th/9304159.}

\lref\maha{J.~Maharana and J.H.~Schwarz, {\it Nucl.~Phys.}~B390 (1993)
3, hep-th/9207016.}

\lref\mastI{J.M.~Maldacena and A.~Strominger, Santa Barbara preprint
UCSBTH-97-02, hep-th/9702015.}

\lref\mast{J.M.~Maldacena and A.~Strominger, \pr D55 (1997) 861, 
hep-th/9609026.}

\lref\mst{J.M.~Maldacena and A.~Strominger, {\it Phys.~Rev.~Lett.}~77
(1996) 428, hep-th/9603060.}

\lref\ms{J.M.~Maldacena and L.~Susskind, 
{\it Nucl.~Phys.}~B475 (1996) 679, hep-th/9604042.}

\lref\myers{C. Johnson, R. Khuri and R. Myers, 
{\it Phys. Lett.} B378 (1996) 78, hep-th/9603061.} 

\lref\schw{J.H.~Schwarz, {\it Nucl.~Phys.}~B226 (1983) 269.}

\lref\sv{A.~Strominger and C.~Vafa, {\it Phys.~Lett.}~B379 (1996) 99,
hep-th/9601029.}


\newsec{Introduction}

Effective string models of $D=5$ black holes with three $U(1)$ charges
\refs{\sv,\ATT,\cm,\hms} 
and of $D=4$ black holes with four $U(1)$ charges \refs{\CY,\CTT}
are being actively
explored in the current literature. In the $D=5$ case the effective string 
models the dynamics of the intersection of D-branes \refs{\sv,\cm,\ms}, 
while in the
$D=4$ case -- that of triply intersecting 5-branes of M-theory \KT.
The initial success of the models was in reproducing the 
Bekenstein-Hawking entropy of 
black holes \refs{\sv,\cm,\hms,\myers,\mst,\KT}, but
more recently the emphasis
has shifted to more dynamical comparisons -- those involving emission and
absorption rates of massless quanta. For minimally coupled scalar fields
such calculations were carried out in 
\refs{\cm,\dmw,\dm,\dmII,\GK,\mast,\GKtwo,\km,\hawk,\dowk}. 
Remarkably, it was found that the energy-dependence of the semiclassical
absorption cross-sections (the so-called greybody factors)
are correctly reproduced by effective string
calculations at sufficiently low energies \refs{\mast,\GKtwo}.
This success has been attributed to the validity of the moduli space 
approximation \juanI. 

An important issue is whether the effective string continues to be
a good description beyond this regime. A good test for this is provided
by the fixed scalars \refs{\fkk,\gkk}, 
whose non-minimal couplings to the gauge fields
render their greybody factors different from those of the 
minimally coupled scalars \refs{\kr,\CGKT}.\foot{Another test is
to compare the absorption of minimally coupled scalars in higher partial
waves, which appears to agree up to normalization factors 
\refs{\mastI,\gunp}.}
In \CGKT\ the effective string explanation of the new greybody factors
was traced to the fact that the leading coupling of fixed scalars is
to operators of dimension higher than $(1,1)$. One of the $D=5$
fixed scalars, related to the volume of the internal $T^4$ over which
the 5-branes are wrapped, and called $\nu$ in \CGKT, was found to couple to
an operator of dimension $(2,2)$. The subsequent string calculation
of the absorption cross-section yielded precise agreement with
the semi-classical greybody factor \CGKT.

However, an important technical obstacle, which arises
in the classical supergravity, put a restriction on the
range of comparisons that could be carried out in \CGKT. 
For general 1-brane and 5-brane charges, $Q$ and $P$, the fluctuations
of the two fixed scalar fields, $\nu$ and $\lambda$,  mix with each other
and also with the fluctuations of the metric. For this reason,
the comparison carried out in 
\CGKT\ was limited to the simplest case of $P=Q$, where only $\lambda$
mixes with gravity while $\nu$ is unmixed.
In this paper we overcome this obstacle and disentangle the fixed scalar
equations for $P\neq Q$.
The resulting pair of equations are remarkably simple and are
very similar to the fixed scalar equation derived in \CGKT.
In fact,
the greybody factors that follow from them are both proportional to the
greybody factor calculated in \CGKT. This turns out to disagree with the
effective string action derived in
\CGKT. Even for $P=Q$ the $\lambda$ greybody factor
is not in agreement, while for $P\neq Q$ neither greybody factor 
appears to agree. 
The disagreement is caused by the appearance of
chiral operators with dimensions $(3,1)$ and $(1,3)$ in the
effective string action.

The organization of the paper is as follows. In section 2 we discuss
the simplest situation where a fixed scalar 
arises: the $D=4$ example, which was studied in \kr\ for
equal charges. We show how to decouple the fixed scalar fluctuations
from gravity even for unequal charges and derive the resulting equation.
In section 3 we proceed to the more complicated $D=5$ example, whose
advantage is that it can be directly compared with the effective string.
We derive two decoupled equations for appropriate mixtures of the
original fields $\nu$ and $\lambda$. Comparison of the resulting greybody
factors with those that follow from the effective string calculations
is presented in section 4.
We conclude in section 5.

\newsec{The $D=4$ case}

First, we consider the simpler case of 
the extremal black hole in $D=4$ with two $U(1)$ 
charges and one fixed scalar (the dilaton) \refs{\kaaa,\kr}. 
The action to which this black hole is a solution is
$$ S = \int d^4x \sqrt {-g}
[R-2(\partial_{\mu }\phi )^2-e^{-2\phi }
F_{\mu \nu }^2-e^{2\phi }G_{\mu \nu }^2]\ .$$
The resulting equations of motion are:
\eqn\rfield{\partial_{\mu }(\sqrt {-g}e^{-2\phi }
F^{\mu \nu })=\partial_{\mu }(\sqrt {-g}e^{2\phi }G^{\mu \nu })=0\ ,}
\eqn\rfix{(\partial_{\mu }\phi )^2+{1 \over 2}
e^{-2\phi }F^2-{1 \over 2}e^{2\phi }G^2=0 \ ,}
\eqn\rgrav{R_{\mu \nu }+2\partial_{\mu }\phi 
\partial_{\nu }\phi+e^{-2\phi }
(2F_{\mu \lambda }F_{\nu \delta }
g^{\lambda \delta }-{1 \over 2}
g_{\mu \nu }F^2)+e^{2\phi }(2G_{\mu \lambda }
G_{\nu \delta }g^{\lambda \delta }-{1 \over 2}g_{\mu \nu }G^2)=0\ .} 

We are looking for spherically symmetric 
perturbations, so we will take the metric to be of the form
\eqn\rmetric{ds_4^2 = -e^{2A}dt^2 + e^{2B}dr^2 + r^2e^{-2U}d\Omega _2^2\ ,}
where $A$, $B$ and $U$ depend on $r$ and $t$ only. 
The gauge invariance present in the problem will later allow 
us to specify the precise form of the
function $U$.

Since we are interested in solutions with fixed 
charges, we first solve for the $U(1)$ fields. From \rfield\ we have
$$\partial_r(r^2e^{A+B-2U-2\phi }F^{rt})=
\partial_r(r^2e^{A+B-2U+2\phi }G^{rt})=0 \ . $$
Let the $F$-field carry charge $Q$ and the $G$-field 
carry charge $P$. Then we get
\eqn\fsol{F^{rt} = {Qe^{-A-B+2U+2\phi } \over r^2}\ , 
\qquad F^2 = {-2Q^2 \over r^4}e^{4U+4\phi }\ ,}
and
\eqn\gsol{G^{rt} = {Pe^{-A-B+2U-2\phi } \over r^2}\ , 
\qquad G^2 = {-2P^2 \over r^4}e^{4U-4\phi }\ .}
Substituting \fsol\ and \gsol\ into \rfix\ we obtain

\eqn\nrfix{-\partial_t^2\phi +{1 \over {\sqrt -g}}
\partial_r ({\sqrt -g}\partial_r\phi) -{Q^2 \over r^4}
e^{4U+2\phi }+{P^2 \over r^4}e^{4U-2\phi } = 0\ .}

We are interested in deriving the fluctuation 
equation for $\phi $ around the static black hole solution. 
This solution is \kaaa:
\eqn\rmstat{ds_0^2 = -e^{2U}dt^2+e^{-2U}(dr^2+r^2d\Omega ^2)\ ,}
\eqn\rfstat{
\eqalign{& e^{-2U}=H_1H_2\ , \qquad e^{2\phi_0 }=
{H_2\over H_1}\ ,\cr & 
F={1 \over {\sqrt 2}}dH_1^{-1}\wedge dt\ , \qquad 
G={1 \over {\sqrt 2}}dH_2^{-1}\wedge dt\ ,\cr } }
\eqn\rhstat{H_1=1+{{\sqrt 2}Q \over r}\ , 
\qquad H_2=1+{{\sqrt 2}P \over r}\ .}

We now let both the metric and 
$\phi $ fluctuate, taking the metric to be of the form \rmetric, 
$$A = U(r)  +\delta A (r, t)
\ ,\qquad  B = -U(r) + \delta B(r,t)\ , 
\qquad \phi = \phi_0(r) + \delta \phi(r,t) \ .$$
That is, we keep the angular part of the metric fixed, 
which we can achieve by a gauge transformation. 
Note that the fluctuations are functions of 
$r$ and $t$ only. For the $\phi $ field this means that 
we consider only the $l=0$ partial wave. 
At low frequencies, this gives the dominant contribution to the 
absorption cross-section.

We will solve the equations of motion to first order 
in the fluctuations. First, since we are keeping the charges fixed, 
the expressions for the $U(1)$ fields are as above. 
We now turn to the gravity equations. 
The `$rt$' component of the Ricci tensor is 
$$R_{rt} = -2r^{-1}{\dot B}(1-rU^{\prime})\ ,$$
and consequently the `$rt$' equation is
$$ -2r^{-1}{\dot B}(1-rU^{\prime})+2 {\dot \phi}
\phi^{\prime} = 0\ .$$
Taking the variation, and remembering that $\phi_0$ 
is time-independent, we obtain
$$\delta {\dot B} = {r\phi_0^{\prime} 
\over 1-rU^{\prime}}\delta {\dot \phi }\ .$$
This may be integrated to give
\eqn\rrt{\delta B = {r\phi_0^{\prime}
\over 1-rU^{\prime}}\delta \phi \ .}

Next, we use the angular Einstein equation 
(the `$\theta \theta $' and `$\phi \phi$' 
components yield the same equation):
$$\eqalign{&-1-re^{-2U-2B}[(B^{\prime}+U^{\prime}-
A^{\prime})(1-rU^{\prime})+U^{\prime}+
rU^{\prime \prime}-r^{-1}(1-rU^{\prime})^2]\cr & -
{1 \over 2}e^{-2\phi }g_{\theta \theta }
F^2-{1 \over 2}e^{2\phi }g_{\theta \theta }G^2 = 0\ . \cr } $$
Inserting the expressions for the fields and 
taking the variation, we obtain
\eqn\rff{\delta A^{\prime}-\delta B^{\prime} = 
{-2\delta B \over r(1-rU^{\prime})}
(r^2U^{\prime \prime}+2rU^{\prime}-1)-
{2 e^{2U} \over r^3(1-rU^{\prime})} (Q^2 e^{2\phi _0} - P^2
e^{-2\phi_0}) 
\delta \phi \ .}
\rrt\ and \rff\ will be sufficient to decouple the 
fixed scalar fluctuations from the gravity fluctuations. 

We now turn to the fixed scalar equation \nrfix. Taking 
the variation, and considering fluctuations of the form
$e^{i\omega t}\delta \phi (r)$, we get
$$ \eqalign{ & r^{-2}\partial_r (r^2\partial_r\delta \phi) +
\omega ^2 e^{-4U}\delta \phi- 2\delta Br^{-2}
\partial_r (r^2\partial_r\phi _0) -
\phi _0^{\prime}(\delta B^{\prime}-
\delta A^{\prime})\cr &
-\left ({2Q^2 \over r^4}e^{2U+2\phi _0}+
{2P^2 \over r^4}e^{2U-2\phi _0}\right )\delta \phi = 0\ .\cr }
 $$
Substituting  for 
$\delta A^{\prime}-\delta B^{\prime}$ from \rff\ and 
for $\delta B$ from \rrt, as well as for $U$ and 
$\phi_0$ from \rfstat, we find that the dilaton fluctuations obey 
the following simple equation:
\eqn\fourresult{ \left [ r^{-2}\partial_rr^2\partial_r + 
\omega ^2(H_1H_2)^2-{4(P+Q)^2 \over r^2(\sqrt {2}P+
\sqrt {2}Q+2r)^2} \right ]\delta \phi = 0 \ . }
This is essentially the same equation as 
that obtained in \kr\ for the special case $P=Q$, 
but with the charge $P$ in the potential replaced 
by the average of the two charges, $(P+Q)/2$. 
We see that the unmixing of the gravitational fluctuations for unequal
charges results in the same type of equations as found for equal charges,
but with new parameters. Remarkably, this phenomenon occurs also
for the $D=5$ black hole which we discuss next.

\newsec{The $D=5$ case}

In this section we address our main goal: decoupling 
the fixed scalar fluctuations for the five-dimensional 
black hole with three unequal charges.
The action to which this black hole is a solution is \CGKT
\eqn\fiveact{ S = {1 \over 2\kappa_5^2} \int d^5x {\sqrt -g}
\left [R-{4 \over 3}(\partial_{\mu} \lambda)^2 - 
4(\partial_{\mu} \nu)^2-{1 \over 4}e^{{8 \over 3}\lambda } 
F_{\mu \nu }^{(K)2}-{1 \over 4}e^{-{4 \over 3} 
\lambda +4\nu }F_{\mu \nu }^2-{1 \over 4}
e^{-{4 \over 3} \lambda -4\nu }H_{\mu \nu }^2 \right ]\ . }
We omit the dilaton $\phi$, which in this case is a 
minimally coupled scalar, since it can be set 
to $0$ in what follows. 

We will now proceed in 
precise analogy with the four-dimensional case, 
with the only differences lying in technical details.
We take the metric to be of the form
\eqn\vmetric{ds_5^2 = -e^{2A}dt^2 + 
e^{2B}dr^2 +r^2e^{-2U}d\Omega _3^2\ ,}
where $A$, $B$, and $U$ are functions of $r$ and $t$ only.
The equations obtained by varying with respect to the $U(1)$ fields are:
\eqn\vfield{\partial_{\mu }(\sqrt {-g}e^{{8\over 3}\lambda }
F^{(K)\mu \nu }) = \partial_{\mu }
(\sqrt {-g}e^{-{4\over 3}\lambda +4\nu }F^{\mu \nu }) 
= \partial_{\mu }(\sqrt {-g}
e^{-{4\over 3}\lambda-4\nu }H^{\mu \nu }) = 0\ .}
The $U(1)$ fields will carry fixed charges 
$Q_K$, $Q$ and $P$. Solving \vfield\ we get
\eqn\vfksol{F^{(K)rt} = {2Q_K \over r^3}
e^{-A-B+3U-{8\over 3}\lambda }\ , \qquad 
F^{(K)2} = -{8Q_K^2 \over r^6}e^{6U-{16\over 3}\lambda }\ ,}
\eqn\vfsol{F^{rt} = {2Q \over r^3}
e^{-A-B+3U+{4\over 3} \lambda -4\nu } \ , \qquad
F^2 = -{8Q^2 \over r^6}e^{6U+{8\over 3}\lambda -8\nu }\ ,}
\eqn\vhsol{H^{rt} = {2P \over r^3}
e^{-A-B+3U+{4\over 3}\lambda +4\nu }\ , \qquad 
H^2 = -{8P^2 \over r^6}e^{6U+ {8\over 3}\lambda +8\nu } \ .}
By varying \fiveact\ with respect to the metric 
we get the following equations of motion,
$$ R_{\mu \nu } + {4 \over 3}\partial_ {\mu }
\lambda \partial_ {\nu }\lambda - 4\partial_ {\mu } 
\nu \partial_ {\nu }\nu +e^{{8 \over 3}\lambda }
\left ({1 \over 2}F^{(K)}_{\mu \lambda }F^{(K)}_{\nu \delta }-
{1 \over 8}g_{\mu \nu }F^{(K)2} \right )+e^{-{4 \over 3} 
\lambda +4\nu }\left ({1 \over 2}F_{\mu \lambda }
F_{\nu \delta }-{1 \over 8}g_{\mu \nu }F^2 \right ) $$
\eqn\vgrav{+e^{-{4 \over 3} \lambda -4\nu }
\left ({1 \over 2}H_{\mu \lambda }H_{\nu \delta } -
{1 \over 8}g_{\mu \nu }H^2 \right )=0\ .}
The equations for the fixed scalars are, 
after inserting the metric and fields from \vmetric, \vfksol, \vfsol\ 
and \vhsol,
$$\partial_t \left ({8 \over 3}r^3e^{-3U+B-A}{\dot \lambda } \right )-
\partial_r \left ({8 \over 3}r^3e^{-3U+A-B}\lambda^{\prime} \right )-$$
\eqn\vlam{-2r^{-3}e^{A+B+3U}\bigg [{8 \over 3}Q_K^2
e^{-{8 \over 3}\lambda }-{4 \over 3}P^2
e^{{4 \over 3}\lambda +4\nu }-{4 \over 3}Q^2e^{{4 \over 3}\lambda -4\nu }
\bigg ]=0}
and
$$\partial_t(8r^3e^{-3U+B-A}{\dot \nu })-
\partial_r(8r^3e^{-3U+A-B}
\nu^{\prime})-$$
\eqn\vnu{-2r^{-3}e^{A+B+3U}[-4P^2
e^{{4 \over 3}\lambda +4\nu }+
4Q^2e^{{4 \over 3}\lambda -4\nu }]=0\ .}
We are interested in deriving the fixed scalar 
fluctuation equations around the static solution given in
\refs{\ATT,\cm,\CYY,\hms},
$$e^{-2U}=(H_{\Qk }H_{\P }H_{\Q })^{1/3}\ ,
\quad B_0=-U-{1 \over 2}\ln h\ ,\quad A_0=2U+{1 \over 2}\ln h\ ,$$
\eqn\vstat{e^{2\lambda _0}=H_{\Qk }(H_{\Q }H_{\P })^{-1/2}
\ , \qquad e^{4\nu _0}=H_{\Q }H_{\P }^{-1}\ , }
where
$$h=1-{r_0^2\over r^2}\ , 
\qquad H_{\q }=1+{\q \over r^2}\ , 
\qquad \q = \sqrt {q^2+{r_0^4\over 4}}- {r_0^2\over 2}
\ ,\qquad q = Q_K, Q, P\ .$$
Here $r_0 $ is 
the radius of the horizon, i.e. the 
parameter governing the non-extremality of the solution.

We now let the metric and the fixed scalars 
vary, keeping the angular part of the metric 
fixed. Thus we have
$$A=A_0+\delta A\ ,\quad B=B_0+\delta B\ ,\quad 
\lambda = \lambda _0+
\delta \lambda\ ,\quad \nu = \nu _0+\delta \nu\ ,$$
but $U$ is kept fixed. Again, we can do this 
because of the gauge freedom. We allow the 
fluctuations to depend on $t$ and $r$ only, since for 
sufficiently low frequencies the $l=0$ partial wave will give the 
dominant contribution to the absorption cross-section.

To decouple the fixed scalar fluctuations from 
the metric fluctuations, we look, as before, 
at the `$rt$' and the angular Einstein equations. 
The `$rt$' component of the Ricci tensor $R_{\mu \nu }$ is
$$R_{rt} = -3r^{-1}(1-rU^{\prime}){\dot B}\ . $$
The corresponding equation of motion is found from \vgrav\ to be
$$-3r^{-1}(1-rU^{\prime}){\dot B}+
{4 \over 3}{\dot \lambda } 
\lambda ^{\prime} +4{\dot \nu }\nu ^{\prime} =0\ .$$
Taking the variation, we find
$$ \delta {\dot B} = {r \over 3(1-rU^{\prime})}
\left ({4 \over 3}\lambda _0^{\prime}
\delta {\dot \lambda } +4\nu _0^{\prime}\delta {\dot \nu } \right )\ .$$
This is integrated to give
\eqn\vrt{\delta B = {r \over 3(1-rU^{\prime})}
({4 \over 3}\lambda _0^{\prime}\delta 
\lambda +4\nu _0^{\prime}\delta \nu )\ .}
>From \vgrav\ the angular Einstein equation
is found to be
$$-2-e^{-2U-2B}r[(B^{\prime}+U^{\prime}-
A^{\prime})(1-rU^{\prime})+U^{\prime}+
rU^{\prime \prime}-2r^{-1}(1-rU^{\prime})^2]+$$
$$+{2e^{4U} \over 3r^4}[Q_K^2e^{-{8 \over 3}
\lambda }+Q^2e^{{4 \over 3}\lambda -4\nu }+
P^2e^{{4 \over 3}\lambda +4\nu }]=0\ .$$
Taking the variation, we find
$$\delta A^{\prime} - \delta B^{\prime} 
= -{2\delta B \over r(1-rU^{\prime})}
\left [r^2U^{\prime \prime}+3rU^{\prime}-2-
{rh^{\prime}(1-rU^{\prime})\over h} \right ]+
$$
\eqn\vff{
+{2h^{-1}e^{4U} \over 3r^5(1-rU^{\prime})}
\left [{8Q_K^2 \over 3}e^{-{8 \over 3}
\lambda _0}\delta \lambda-Q^2e^{{4 \over 3}
\lambda _0-4\nu _0}({4 \over 3}\delta \lambda -4\delta \nu )
-P^2e^{{4 \over 3}\lambda _0+4\nu _0}
({4 \over 3}\delta \lambda +4\delta \nu ) \right ]\ .}
Again, the relations \vrt\ and \vff\ will suffice to decouple 
the fixed scalar fluctuations from the gravity 
fluctuations. Taking the variations of the fixed 
scalar equations \vlam\ and \vnu\ with frequency $\omega $, 
and using \vrt, \vff\ and \vstat, 
we get the following two coupled equations
\eqn\coupledone{[r^{-3}\partial_rhr^3\partial_r+
\omega ^2h^{-1}H_{\Qk}H_{\Q}H_{\P}+f_1(r)]
\delta \tilde \lambda +\sqrt 3 f_2(r)\delta \nu =0 }
and
\eqn\coupledtwo{ [r^{-3}\partial_rhr^3\partial_r+
\omega^2h^{-1}H_{\Qk}H_{\Q}H_{\P}+f_3(r)]
\delta \nu +\sqrt 3 f_2(r)\delta \tilde \lambda =0 \ ,}
where we have defined 
$$ \delta \lambda = \sqrt 3 \delta \tilde \lambda\ ,
$$
so that the kinetic terms for $\delta \nu$ and $\delta \tilde \lambda$
have the same normalization in the action.
The functions entering the fixed scalar equations have the following
form,
$$
\eqalign{ f_1(r)=&
-{8\over r^2 [\P\Q+\P\Qk+\Q\Qk+2(\P+\Q+\Qk)r^2+3r^4]^2 } 
\times  \big [\P^2\Q^2+\P^2\Qk^2 +\Q^2\Qk^2 \cr  &
+2\P\Q\Qk(\P+\Q+\Qk) +{3\over 2}r_0^2(\P+\Q)(\P\Q+\P\Qk+\Q\Qk) \cr 
& +\big ( (\P\Q+4\Qk^2) ( \P+\Q) +\Qk (\P^2+\Q^2) +
6\Qk\P\Q +6r_0^2(\P\Q+\P\Qk+\Q\Qk)\big )r^2 \cr 
&+\big (\P^2-\P\Q+\Q^2+4\Qk^2+2\Qk\P+2\Qk\Q+ {3\over 2} r_0^2(\P+\Q+4\Qk)
\big )r^4 
\big ]\ , \cr }
$$

$$\eqalign{ f_2(r)=&{8(\Q-\P)\over r^2
[\P\Q+\P\Qk+\Q\Qk+2(\P+\Q+\Qk)r^2+3r^4]^2}
\times \big [-{1\over 2}r_0^2(\P\Q+\P\Qk+\Q\Qk) \cr & +(\P\Q+\P\Qk+\Q\Qk)r^2+
\big (\P+\Q+\Qk+ {3\over 2} r_0^2 \big )r^4 \big ]\ , \cr}$$

$$
\eqalign{ f_3(r)=&
-{8\over r^2 [\P\Q+\P\Qk+\Q\Qk+2(\P+\Q+\Qk)r^2+3r^4]^2 } 
\times  \big [\P^2\Q^2+\P^2\Qk^2 +\Q^2\Qk^2 \cr 
& +2\P\Q\Qk(\P+\Q+\Qk)+ {1\over 2} r_0^2(\P+\Q+4\Qk)(\P\Q+\P\Qk+\Q\Qk) \cr &  
+3\left (\P\Q (\P+\Q)+(\P^2+\Q^2)\Qk+2\P\Q\Qk +
6r_0^2(\P\Q+\P\Qk+\Q\Qk\right )r^2 \cr
&+3\big (\P^2+\P\Q+\Q^2+ {3\over 2} r_0^2(\P+\Q)\big )r^4 \big ] \ . \cr }
$$

Compared to the $D=4$ case we now encounter 
the additional difficulty that the two fixed scalars 
couple to each other. Luckily, however, 
the fixed scalar equations \coupledone\ and \coupledtwo\ 
may be decoupled by a position-independent
rotation of the fields, 
$$\delta \tilde \lambda = 
(\cos \alpha )\phi_++(\sin \alpha )\phi_- \ ,$$
$$\delta \nu = -(\sin \alpha )\phi_++(\cos \alpha )\phi_- \ ,$$
where the rotation angle satisfies
\eqn\rcondition{\tan \alpha- {1\over \tan\alpha} =
{2\over \sqrt 3} {\P+\Q-2 \Qk\over \Q-\P}
\ . } 
Solving this quadratic equation, we find that
$$ \tan \alpha = {1\over \sqrt 3}
{\P+\Q-2 \Qk \pm 2 
\sqrt {\P^2+\Q^2+\Qk^2-\P\Q-\P\Qk-\Q\Qk}\over \Q-\P}\ , $$
which implies that
$$\cos^2\a = {1 \over 2} \pm  {1 \over 4} {\P+\Q-2\Qk \over \sqrt
{\P^2+\Q^2+\Qk^2-\P\Q-\P\Qk-\Q\Qk}}\ . $$
Using this result, we find that
$\phi_{\pm }$ satisfy the following simple equations, 
\eqn\neweqs{ \left [r^{-3}\partial_r h r^3\partial_r+\omega ^2
h^{-1} H_{\Qk}H_{\Q}H_{\P}-{8Q_{\pm }^2
\over r^2 \left (r^2+Q_{\pm }\right )^2}
\left (1+{r_0^2 \over Q_{\pm }} \right ) \right ]\phi_{\pm }=0\ ,}
where we have defined
$$Q_{\pm } = {1 \over 3}(\P+\Q+\Qk\mp 
\sqrt {\P^2+\Q^2+\Qk^2-\P\Q-\P\Qk-\Q\Qk}).$$
Note that these equations are manifestly symmetric under interchange of any
pair of the charges, i.e. U-duality invariant. This is a nice consistency
check on our results.\foot{This check was suggested by A.~Tseytlin.}

Calculation of the absorption cross-sections from
\neweqs\ in the ``dilute gas regime'' 
($r_0^2, Q_K \ll P,Q$) is analogous to that
presented in \CGKT, and we find that the greybody factors are
proportional to the $\nu$ greybody factor found for $Q=P$.
The coefficient of proportionality is a function of $P$ and $Q$.
The absorption cross-sections for $\phi_\pm$ are
\eqn\Sigmapm{
\sigma_\pm =
    {9 \pi^3 P^3 Q^3\over 
64(P+Q\mp \sqrt {P^2-PQ+Q^2})^2}
   {\omega \left (\e^{\omega\over T_H} - 1 \right ) \over
   \left (\e^{\omega\over 2 T_L} - 1\right )
   \left (\e^{\omega\over 2 T_R} - 1\right ) }
   (\omega^2 + 16 \pi^2 T_L^2) (\omega^2 + 16 \pi^2 T_R^2) \ ,
  }
where $T_H$ is the Hawking temperature,
while $T_L$ and $T_R$, which are determined by $r_0$ and the charges
\mast, play the role of the left- and right-moving temperatures
on the effective string. 
In the next section we compare \Sigmapm\
with the results of the effective string model.

\newsec{Comparison of Greybody factors}

The greybody factors one finds from the two equations \neweqs\ in
general disagree with the predictions of the effective string 
action derived in \CGKT.
This even happens for $Q=P$ where there is no
mixing between $\nu$ and $\lambda$. In \CGKT\ agreement was found
for the scalar field $\nu$. However, now that we have derived the
equation for $\lambda$, we will see that for this scalar there is no 
agreement.

Setting $P=Q$ in \Sigmapm,
we find that the classical absorption cross-section for $\lambda$ is
\eqn\SigmaClass{
   \sigma_{\rm abs}(\omega) = { 9\pi^3 P^4\over 64}
   {\omega \left (\e^{\omega\over T_H} - 1 \right ) \over
   \left (\e^{\omega\over 2 T_L} - 1\right )
   \left (\e^{\omega\over 2 T_R} - 1\right ) }
   (\omega^2 + 16 \pi^2 T_L^2) (\omega^2 + 16 \pi^2 T_R^2) \ ,
  }
On the effective string side, the $\lambda$-coupling is \CGKT
\eqn\iise{
-  {T_{\rm eff}\over 8 } \lambda 
[ \del_+ X \del_- X  (  (\del_+ X)^2  +
 (\del_- X)^2)   +   (\del_+ X)^2  (\del_- X)^2] }
plus the fermionic terms required by supersymmetry ($T_{\rm eff}$
is the effective string tension).
The last term is an operator of dimension $(2, 2)$ which also couples
to $\nu$. Its effects were studied in \CGKT, and the resulting contribution
to the cross-section is
\eqn\SigmaDbrane{
   \sigma_1(\omega) = { \pi P^2 \over 1024 T^2_{\rm eff}}
   {\omega \left (\e^{\omega\over T_H} - 1 \right ) \over
   \left (\e^{\omega\over 2 T_L} - 1\right )
   \left (\e^{\omega\over 2 T_R} - 1\right ) }
   (\omega^2 + 16 \pi^2 T_L^2) (\omega^2 + 16 \pi^2 T_R^2) \ ,
  }
which is proportional to \SigmaClass. However, there are additional
contributions to the cross-section arising from the first two operators
in \iise\ which have dimensions $(3, 1)$ and $(1, 3)$. 
These operators give
rise to processes involving 3 left-movers and 1 right-mover or
3 right-movers and 1 left-mover.

Let us consider first the processes with 3 left-movers and 1 right-mover.
Using the methods of \CGKT\ we find that their contribution to the
absorption rate is
\eqn\llr{ \eqalign{&
\sim {\kappa_5^2 L_{\rm eff} \over T^2_{\rm eff} }
{1\over 1- \e^{-{\omega\over 2 T_R}} }
       \int_{-\infty}^\infty d p_1 d p_2 d p_3\,
        \delta\left( p_1 + p_2 +p_3 -{\omega\over 2} \right)
        {p_1 \over 1 - \e^{-{p_1\over T_L}}}
        {p_2 \over 1 - \e^{-{p_2\over T_L}}}
        {p_3 \over 1 - \e^{-{p_3\over T_L}}}
\cr
&
\sim {\kappa_5^2 L_{\rm eff} \over T^2_{\rm eff}}
        {\omega \over
         \left(1- \e^{-{\omega\over 2 T_L}} \right)
         \left(1- \e^{-{\omega\over 2 T_R}} \right)
        }
        \left( \omega^2 + 16 \pi^2 T_L^2 \right) 
        \left( \omega^2 + 32 \pi^2 T_L^2 \right) 
\ .
}}
Processes with 3 left-movers and 1 right-mover make a contribution with
$T_L$ interchanged with $T_R$. Converting the rate to the absorption
cross-section using detailed balance, we find the following
additional contribution on the effective string side,
$$ \eqalign{
\sigma_2(\omega)& \sim {\kappa_5^2 L_{\rm eff} \over T^2_{\rm eff}}
{\omega \left (\e^{\omega\over T_H} - 1 \right ) \over
\left (\e^{\omega\over 2 T_L} - 1\right )
\left (\e^{\omega\over 2 T_R} - 1\right ) } \cr
&\bigg [(\omega^2 + 16 \pi^2 T_L^2 ) (\omega^2 + 32 \pi^2 T_L^2)+
(\omega^2 + 16 \pi^2 T_R^2 ) (\omega^2 + 32 \pi^2 T_R^2) \bigg ]
\ .\cr }
$$
Thus, there is no agreement for the $\lambda$ greybody factors. 
At extremality
(for $T_R=0$)
$\sigma_2$ dominates over $\sigma_1$ for small $\omega$.
This is because $\sigma_1 \sim \omega^2$ while 
$$\sigma_2\rightarrow \sim 
{\kappa_5^2 L_{\rm eff} \over T^2_{\rm eff}} T_L^5
\ .$$
This behavior is in marked disagreement with the fact that
$\sigma_{class} \sim \omega^2$.

We have shown that there is some disagreement between 
the semiclassical and the
effective string cross-sections even for $P=Q$. This could be traced
to the presence of dimension $(1, 3)$ and $(3, 1)$ operators in the 
$\lambda$-coupling, which are coming from the $h_{55}$ part of
$\lambda$. Even more mysterious from the effective string
point of view is the mixing between $\lambda$ and $\nu$ induced
by $P\neq Q$. If one takes the lagrangian derived in \CGKT\ at face
value, then both these mixtures now have coupling to
dimension $(1, 3)$ and $(3, 1)$ operators, which implies disagreement
of the greybody factors for both of them.

\newsec{Conclusions}

Let us summarize our results. The form of the semiclassical greybody 
greybody factors suggests that both $\nu$ and $\lambda$ couple to dimension
$(2,2)$ operators on the effective string. However, the fact that
$\lambda$ contains $h_{55}$ implies that dimension 
$(1, 3)$ and $(3, 1)$ operators are also present in the coupling.
One possibility of restoring agreement between
the supergravity and the effective string
is by finding an overlooked mixing with yet another scalar field.
In fact, a surprising new mixing was recently found for fields
which couple to effective string
operators of dimension $(1, 2)$ and $(2, 1)$ \krt. 
However, we have not been able to find a scalar that mixes
with $\nu$ and $\lambda$, and in general our calculations exhibit a marked
disagreement between the semiclassical and the effective string
greybody factors for these fixed scalars.

We may attempt a different approach: rather than try to derive
the string action, as was done in \CGKT, we could simply guess
the terms that reproduce the semiclassical greybody factors
\Sigmapm. Although this type of modeling is not predictive,
we indeed find that a coupling of the form
\eqn\guess{\int d^2\sigma
\big ( c_+ (P, Q) \phi_+ + c_- (P, Q) \phi_- \big )
T_{++} T_{--}\ ,
}
where $T_{\alpha\beta}$ is the energy-momentum tensor,
can lead to agreement provided that the functions $c_\pm (P,Q)$
are appropriately chosen. It seems difficult, however, to explain the
peculiar form of these functions.

We have shown that the fixed scalars pose a challenge
for the effective string models of black holes.
It will be interesting to see whether there is a way out of this
difficulty.

\newsec{Acknowledgements}

We are grateful to A.~Tseytlin for constructive suggestions and
for many valuable discussions.  This work
was supported in part by DOE grant DE-FG02-91ER40671,
the NSF Presidential Young Investigator Award PHY-9157482, and the
James S.{} McDonnell Foundation grant No.{} 91-48.

\listrefs
\bye